# Possible Analogues of Cognitive Processes in the Patterns of X-ray Variability of the Rapid Burster

Vladimir A. Lefebvre
School of Social Sciences
University of California, Irvine CA 92697
and
Yuri N. Efremov
Sternberg Astronomical Institute
Moscow State University, Moscow

**Abstract**

The bizarre patterns of rapid flashes and bursts in the X-ray source MXB 1730-335 (Rapid Burster) have been puzzling researchers for two decades; especially intriguing are its peculiarities in the time-invariant decay profiles of many type II bursts of the Rapid Burster. We have discovered that they are similar to certain regularities found in cognitive psychology, suggesting an analogy between the activity of the Rapid Burster and the cognitive activity of the human mind.

## 1 Introduction

Twenty-two years after its discovery (Lewin, et al., 1976) the variable X-ray source MXB 1730-335 remains an enigma. The constantly changing modes of its bursts and quasi-periodic oscillations are unique. They display extremely complex and endlessly varying patterns, a state of affairs for which no explanation has been suggested till now. It is one of about 125 low-mass known X-ray binaries, of which Her X-1 was the first to be discovered (Efremov, Sunyaev, and Cherepashchuk, 1974). Approximately 50 of these stars exhibit the bursts (called type I) attributed to thermonuclear flashes on the surface of a neutron stars, in addition to persistent X-ray flux. The unique trait of MXB 1730-335, and the reason why it is called Rapid Burster (RB), is that it produces rapid repetitive X-ray bursts of highly variable yet somewhat regular patterns; the characteristic features of these patterns have been described in almost every new paper on RB. The activity of the RB repeats itself periodically, with intervals of about 6 months between active periods lasting 2-6 weeks.

The extremely complicated variations of the Rapid Burster display intriguing regularities of a kind which have never been found in natural physical systems. In the present paper we would like to draw attention to certain features of the Rapid Burster's X-ray activity which resemble those known to cognitive psychology.

## 2 Double Geometric Progression and Golden Section in the Decay Profiles of RB Bursts

A striking similarity has been found (Tawara, et al., 1985). in the forms of descending parts in different bursts of the type II. These decay structures show a few humps, and if the duration of a burst is normalized with the interval between two peaks in the decay profile, the structure of the latter transforms into a time-scale invariant (Tawara, et al., 1985). It is clear that the decay profile is more complex than a simple damping oscillation: this profile can be described using a "double" geometric progression (see Fig.1).

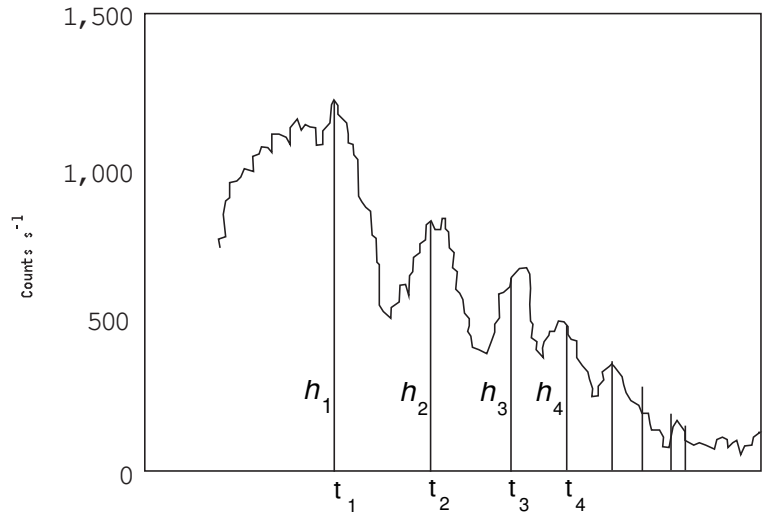

$$h_{2k+1} = h_1 \alpha^k, \quad h_{2k+2} = h_2 \alpha^k, \quad \frac{h_2}{h_1} = \beta$$

$$(k = 0, 1, 2 \ldots)$$

Figure 1: The invariant profile (Tawara, et al., 1985) of the type II bursts in 1984.

The time intervals between even-numbered peaks and those between odd-numbered peaks, respectively, compose geometric progressions (Tawara, et al., 1985) with the same common ratio of approximately $\alpha = 0.57$; this was found both in the observations of 1983 and in those of 1984. The heights of the odd-numbered peaks also form a geometric progression (Tawara, et al., 1985) with a common ratio of $\alpha = 0.57$, and the heights of the even-numbered peaks form a geometrical progression, again with a common ratio of $\alpha = 0.57$. This value is repeated in the observation sets from both 1983 and 1984. The ratio of the second and first peak heights was found to be $\beta = 0.81$ for 1983 and $\beta = 0.77$ for 1984.

The same rule applied both to time intervals and to the heights of the peaks means that the heights of the successive peaks decrease linearly with time. All this implies the presence of some complex flow-control mechanism, if it is assumed that the luminosity of a burst is determined by the instantaneous accretion rate (Tawara, et al., 1985). "One may speculate that the structures are already prepared in the accretion disk (for example, regular concentric rings similar to Saturn rings)," (Tawara et al. 1985, p.547.)

Another group of researchers (Tan et al., 1991) confirmed only partially the findings by Tawara et al. (1985), but they found another intriguing rule in the decay profile of the RB type II bursts. Namely, for bursts lasting less than 25 sec, if two significant peaks are observed in the power spectrum, the ratio of their centroids frequencies is approximately constant and equal to 1.59. This value is averaged from all the sets, including those with a standard error of as much as 0.19. It follows from Table 2 (Tan et al.,1991) that, in sets of observations which give this ratio with a standard error equal to or less than 0.02, its average value is 1.61. We found that this ratio is very close to the golden section ratio $F = (\sqrt{5} + 1)/2 = 1.618....$

## 3 Formal Model of the Subject with Self-Reflexion

In addition to their conjunction in these recent development of astronomical research, the double geometric progression and the golden section have also appeared together within the formal model of the subject capable of carrying out consecutive acts of self-reflexion (Lefebvre, 1980; 1995). Certain of this model's predictions having to do with binary choice have been confirmed experimentally (Adams-Webber, 1997). Several years ago it was found that, from the formal point of view, this model, analogous to a simple construction, is based on only the first and second laws of thermodynamics (Lefebvre, 1995; 1997). The possibility of giving a very general description of the cognitive process of consecutive acts of reflexion is an argument in favor of the hypothesis that such a process can take place in systems with an essentially different physical basis (Lefebvre, 1997). This description is presented as a chain of heat reservoirs with heat engines between them (see Fig.2).

The temperatures of the reservoirs form a decreasing geometrical progression. Every engine is located between two neighboring reservoirs; it works by taking heat from the warmer reservoir and giving it to the cooler one. Each consecutive engine takes from its hot reservoir the heat yielded into it by the preceding engine and produces work equal to the lost available work of the preceding engine. In other words, engine 1 produces some work, but since it is not necessary a reversible engine, part of the work that theoretically could be

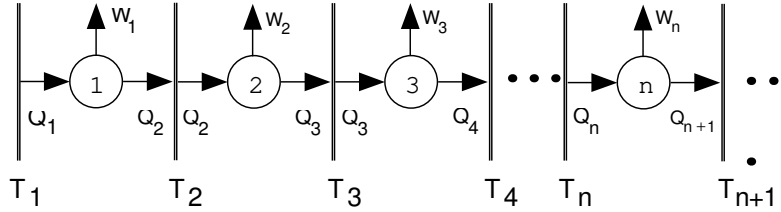

$$\frac{Q_{n+1}}{T_n} = \frac{Q_2}{T_1}, \qquad W_{n+1} = T_{n+1}\left(\frac{Q_{n+1}}{T_{n+1}} - \frac{Q_n}{T_n}\right)$$

Figure 2: A sequence of heat engines. Engine $n$ receives heat $Q_n$ from the reservoir with temperature $T_n$, yields heat $Q_{n+1}$ into the reservoir with temperature $T_{n+1}$, and produces work $W_n = Q_n - Q_{n+1}$.

produced has been lost and the corresponding energy is yielded into the cool reservoir. Engine 2 takes from this reservoir the heat exactly equal to what engine 1 yielded into it and produces work compensating for the "inefficiency" of the first engine. Therefore, engine 2 produces just as much work as engine 1 underproduced. In doing so, engine 2 produces less work than its theoretical maximum, and so engine 3 compensates for the "inefficiency" of engine 2 in the same way that engine 2 compensates for engine 1, and so forth. We denote the engines' respective efficiencies as $x_1, x_2, x_3, \ldots$.

It turns out that such a sequence of engines has certain formal peculiarities (Lefebvre, 1995). First of all, the efficiency of each engine with an odd number is equal to the efficiency of engine 1, $x_1$, and that of those with even numbers is equal to $x_2$. Therefore, the efficiency of the engines forms an alternating sequence: $x_1, x_2, x_1, x_2, \ldots$.

Second, the work produced by the engines forms a double geometrical progression. The work of the odd-numbered engines ($n = 2k+1$) is equal to

$$W_{2k+1} = W_1(T2/T1)^k \tag{1}$$

and the work of the even-numbered engines ($n = 2k+2$) is equal to

$$W_{2k+2} = W_2(T2/T1)^k, \tag{2}$$

where $T_1$ and $T_2$ are the temperatures of the 1st and 2nd reservoirs respectively and $k = 0, 1, 2\ldots$.

Within the framework of the psychological model of reflexion, engine 1 is a theoretical analogue to the subject. Engine 2 is an analogue to the image of the self. The subsequent engines represent the hierarchy of images of the self arising in consecutive acts of self-reflexion. Correspondingly, the work performed by engine 1 is interpreted as emotions experienced by the subject. The work performed by the subsequent engines is interpreted as a hierarchy of cognitive representations of the self's emotions.

The value of $x_1$ corresponds to a stimulus acting on the subject, and that of $x_2$ to the cognitive representation of that stimulus.

Certain stimuli have no inherent measure of intensity in relation to the human perceptual system under the given conditions. In situations of choice involving stimuli of this kind, as has been shown (Lefebvre, 1985), the parameters $x_1$ and $x_2$ take on the values

$$x_1 = (1/2), x_2 = (1/F) = 0.618...$$

The model predicts that a subject in this state will not choose between two alternatives with random, that is equal probabilities; rather the probability of choosing one of the alternatives will be 0.618, that of choosing the other, 0.372. This theoretical finding allows us to explain various phenomena of asymmetry in human choice (Lefebvre, 1985; 1992; 1995; 1997; Benjafield and Adams-Webber, 1976; Adams-Webber, 1987).

## 4 Interpretation of the Activity of the Rapid Burster as Analogous to Cognitive Processes

Let us now put the sequence of peaks in Fig.1 into correspondence with the sequence of engines in Fig.2: the first engine corresponds to the first peak, the second engine corresponds to the second one, and so on. Then the sequence of their heights $h_1, h_2, h_3$ corresponds to the sequence of quantities work performed $W_1, W_2, W_3, ...$ . The link between the sequence of engines and parameters $\alpha$ and $\beta$ is given by the equations $T_2/T_1 = \alpha$ and $W_2/W_1 = \beta$. The ratios between $W_1$, $W_2$, and $W_3$ are proportional to the values of

$$x_1, (1-x_1)x_2, (1-x_1)(1-x_2)x_1. \tag{3}$$

The values of $x_1$ and $x_2$ are linked with $\alpha$ and $\beta$ by the following system of equations:

$$(1-x_1)(1-x_2) = \alpha,$$
$$((1-x_1)x_2)/x_1 = \beta. \tag{4}$$

In the observations for 1983, $\alpha = 0.57$ and $\beta = 0.81$. By solving system (4) we obtain $x_1 = 0.238$ and $x_2 = 0.252$. In the observations for 1984, $\alpha = 0.57$ and $\beta = 0.77$; hence, $x_1 = 0.243$ and $x_2 = 0.247$. These data could be interpreted as evidence that, in the observation period 1983-84, the values of $x_1$ and $x_2$ were approaching a magnitude of 0.25. With $x_1 = x_2 = 0.25$, the exact values of parameters observed are $\alpha = (9/16) = 0.5625$ and $\beta = (3/4) = 0.75$. Note that for $x_1 = x_2$, the double geometrical progression turns into a regular one with a common ratio $\beta$. In this case $\alpha = \beta^2$. The equality of efficiencies of engines at odd and even positions in the sequence suggests an important interpretation within the framework of the cognitive model. It implies that a subject which is modeled by the engine sequence has correct self-reflexion (Lefebvre, 1997).

What is the meaning of the fact that the relations between peak heights correspond to the relation between measures of work produced by heat engines connected in a chain? We can answer this question in two fundamentally different ways. The first is that the peaks are generated directly by some physical process whose working is analogous to that of the chain of heat engines. The second is that the decay profile is a special signal carrying information about the cognitive process connected both with the double geometric progression and with golden section. In the second case the physical mechanism generating a decay profile might be fundamentally different. The second answer seems more probable, since the peak heights are not a direct reflection of the energy given off by RB.

## 5 Discussion

Our conclusion concerning the similarity of the parameters of some processes displayed by Rapid Burster to those of certain phenomena studied in cognitive psychology opens the possibility of an intrinsic similarity between the RB activity and human cognition. The traits indicated are, to be sure, not the only ones. Every new set of observations of the RB uncovers new enigmatic features (Lubin et al., 1992). For example, Dottani et al. (1990) found 5 Hz quasi-periodic oscillations during type II bursts, which are very strong during the initial peak of the burst, absent in the second peak, and strong again in the third peak. These oscillations can be described as periodical ones, with the periods varying up to 25% during a burst; the changes of periods are sometimes periodical themselves. The authors (Dottani et al., 1990) write that the oscillations "know" what their frequency is; they also note that, although some speculation on the properties of the reservoir may lead to an explanation of the "odd-even" behavior of the quasi-periodic oscillations, "it contributes little to answering the important question as to what it is that makes this source so unique." As Kawai et al. (1990) note, the time invariant of the profiles of type II bursts indicates that, at the beginning of the burst, "the system knows how large the burst is going to be." Analogous phenomena are experienced in human cognitive activity, as for instance in spontaneous speech: the beginning of a given sentence's structure depends on the final structure of the sentence taken as a whole.

It is worth noting that the RB is located in the highly reddened globular cluster Liller 1 at a distance of about 10 kpc (Rutledge et al., (1995). Given the uncertainty of the cluster distance, it may be located rather close to the center of the Galaxy. In any event, the occurrence of the RB in a globular cluster implies that the age of its progenitor is about 15 Gyrs.

A radio counterpart to the RB was recently found between 4-5.6 sigma away from the X-ray position (Rutledge et al., 1997). The radio on/off behavior was found to be correlated with the X-ray on/off behavior; the authors (Rutledge et al., 1997) conclude that the accidental probability of an unrelated background source is thus only 1.6%. We found that the coordinates of the radio source (as given in the IAU Circ. # 6813, 1998) are precisely the same as those for the center of the cluster Liller 1, whereas the RB itself is offset from the center by $\sim 8$". This makes the situation even more fascinating. If the same rapid bursts are found coming from a radio source which is a plausible counterpart of the RB, this could well be regarded as evidence for the artificial origins of one or both of them.

# 6  Conclusions

Whatever the peculiarities of the RB are, at times (probably in the beginning of the turn-on state) it can behave like other normal transient low-mass X-ray bursters, generating persistent emissions and type I (thermonuclear) bursts, as Tan et al. (1991) note. They conclude that the RB probably differs from other bursters by just one ingredient in the neutron star binary system.

May this ingredient have something to do with cognition in a non-biological system formed by plasma and governed by a magnetic field? As we noted earlier, some of those who have studied the RB use unusual wording to describe its bizarre and constantly varied behavior, phrases such as "it knows," "it anticipates". Thus, the idea of regarding the Rapid Burster as potentially a "cognitive object" (as has been noted (Lefebvre, 1997)) or even cosmic civilization seems to have come of age.

The possibility of using the X-ray band to search for extraterrestrial intelligence (SETI) has been long discussed (Corbet, 1997). As Shvartsman (1981) has noted, X-ray and optical ranges have an advantage over radio range in the area of maximal bit rate, which could be presumed necessary for communication with advanced civilizations. Note also that the X-ray band of the spectrum, unlike the optical band, has the important property of not suffering from light absorption. The possibility that the X-ray band might be used by an extraterrestrial intelligence is discussed by Fabian (1977), who suggests that X-ray pulses can be generated by projecting something like rocks onto the surfaces of neutron stars. These propositions were made in connection precisely with the discovery of X-ray bursters. As Fabian (1977) notes, a mass of $10^{13}$ kg could produce an X-ray burst of $10^{36}$ erg, which would be detectable in all parts of the Galaxy. He also notes that another way to modulate X-ray luminosity is with the aid of an orbiting metal screen of about one million km in size.

Corbet (1997) has recently suggested that cosmic civilizations could use quasi-periodic oscillations and bursts of X-ray binaries to transmit information. The preceding analysis should indicate that this hypothesis deserves the most serious consideration.

# 7 Acknowledgement